\documentclass[12pt]{article}
\usepackage{epsfig}

\def \sect #1 {\setcounter{equation} 0\section{#1}}
\def \be  {\begin{equation}}
\def \ee  {\end{equation}}
\def \ba  {\begin{eqnarray}}
\def \ea  {\end{eqnarray}}
\def \baa {\begin{eqnarray*}}
\def \eaa {\end{eqnarray*}}
\def \bb  {\begin{thebibliography}}
\def \eb  {\end{thebibliography}}

\def \lab #1 {\label{#1}}

\def \fracs #1#2 {\mbox{\small $\frac{#1}{#2}$}}

\def \bin #1#2 {{\left({#1}\atop{#2}\right)}}
\def \as {\relax\ifmmode\alpha_s\else{$\alpha_s${ }}\fi}

\def \al #1 {\frac {\as({#1})}{\pi} }
\def \ds #1 {\ooalign{$\hfil/\hfil$\crcr$#1$}}

\newcommand \bea{\begin{eqnarray}}
\newcommand \eea{\end{eqnarray}}

\def\hepph  #1 {{\tt hep-ph/#1}}

\begin{document}

\begin{flushright}
YITP-SB-05-01\\
\today
\end{flushright}

\vbox{\vskip .5 in}

\begin{center}
{\Large \bf Fragmentation, Factorization and Infrared Poles\\
\medskip

in  Heavy Quarkonium Production}

\vbox{\vskip 0.25 in}

{\large Gouranga C.\ Nayak$^a$, Jian-Wei Qiu$^b$ and George Sterman$^a$}

\vbox{\vskip 0.25 in}

{\it {}$^a$C.N.\ Yang Institute for Theoretical Physics,
Stony Brook University, SUNY\\
Stony Brook, New York 11794-3840, U.S.A.}

\bigskip

{\it {}$^b$Department of Physics and Astronomy,
Iowa State University\\
Ames, IA 50011, U.S.A.}

\end{center}

\begin{abstract}
We explore the role of soft gluon exchange
in heavy quarkonium production at large
transverse momentum.  We find
uncanceled infrared poles at NNLO
that are not matched by conventional NRQCD matrix elements.
We show, however, that gauge invariance
and factorization require that conventional NRQCD
production operators be modified to include
nonabelian phases or Wilson lines.
With appropriately modified operators, factorization
is restored at NNLO.
We also argue that, in the absence of
special cancellations,
infrared poles at yet higher orders
may require the inclusion of
additional nonlocal  operators, not present in the
NRQCD expansion in relative veloctiy.
\end{abstract}

\bigskip


\section{Introduction}

The production of
heavy quarkonium offers a unique perspective into the process
of hadronization, because the creation of the relevant valence partons,
the heavy quarks, is essentially perturbative.
Quarkonium production and decay have
  been the subject of a vast theoretical literature and of intensive 
experimental study,
in which the effective field theory Nonrelativistic QCD
(NRQCD) \cite{bodwin94} has played a guiding role.
NRQCD offers a systematic formalism to separate
dynamics at the perturbative mass
scale of the heavy quarks
from nonperturbative dynamics, through an expansion
in relative velocity within the pair forming the
bound state.  In NRQCD, the description of the relevant nonperturbative 
dynamics
is reduced to the determination of a limited number of QCD
matrix elements, accessible from experiment and, in principle,
lattice computation.  A characteristic feature of the application of 
NRQCD
to production processes is the indispensible role of
color octet matrix elements, which describe
the nonperturbative transition of quark pairs in
adjoint representation into quarkonia through soft gluon
emission.

An  early success of
NRQCD was to provide a  framework for
the striking Tevatron Run I data
on high-$p_T$  heavy quarkonium production \cite{teva},
and it has been extensively applied to heavy quarkonia
in both collider and fixed target experiments.
  A wide-ranging review
of theory and experiment for quarkonium production and decay has  been
given recently in Ref.\ \cite{review}.
  Much of the analysis has been
based on a  factorization formalism proposed in \cite{bodwin94},
which offers a systematic procedure for the application
of NRQCD to quarkonium production.
  It is fair to say, however, that, in contrast
to quarkonium decay, fully convincing arguments
have not yet been given for NRQCD factorization
as applied to high-$p_T$ production processes
\cite{review,bodwin03}.  This omission may or may not be related
to the current lack of confirmation for its predictions on quarkonium
polarization at high $p_T$ \cite{polar}.

In this paper, we summarize progress toward the derivation of
an appropriate factorization formalism
for high-$p_T$ quarkonia, illustrating our considerations
with results on infrared emission at next-to-next-to-leading order 
(NNLO).
At NNLO we find infrared divergences that do not fall precisely into
the pattern suggested in Ref.\ \cite{bodwin94}.
These divergences may, however, be incorporated into
color octet matrix elements by a technical redefinition that makes the 
latter
gauge invariant.  It is not clear whether this pattern extends
beyond NNLO,  and we conclude
that NRQCD factorization  must be examined further for production 
processes.
In any case, all our results are consistent with the
factorization of evolution logarithms in the ratio of
momentum transfer to quark mass from
nonperturbative matrix elements \cite{bodwin94}.

In the results presented below,
the relevant infrared divergence
is proportional to $v^2$, where
$\vec{v}$ is the relative velocity of the heavy pair
in the quarkonium rest frame.  The rotational invariance
of this result (in the quarkonium rest frame)
makes it possible to match the long-distance behavior of
an arbitrary cross section to an
octet matrix element in a manner that does not
depend on the directions of energetic final-state gluons.  In other 
words,
we may factorize the perturbative long-distance
contributions from the short-distance
cross section, and replace them with a universal
nonperturbative matrix element that has the same perturbative
long-distance behavior, just as proposed
in \cite{bodwin94} and extended in \cite{braaten97}.    We begin our 
discussion with a brief
review of NRQCD factorization at high transverse momentum.

\section{NRQCD in Fragmentation}

We discuss for
definiteness the production of the $J/\psi$
and related heavy quarkonium states $H$ in leptonic or hadronic 
collisions,
$A+B\rightarrow H(p_T)+X$.  To leading power in $m_c/p_T$,
which we assume to be a small parameter,
production proceeds through gluon fragmentation.  According
to conventional factorization theorems \cite{1pIfact}, we have (keeping 
only the gluon)
\begin{equation}
d\sigma_{A+B\to H+X}(p_T) = d\hat\sigma_{A+B\to g+X}(p_T/z\mu) \otimes
D_{H/g}(z,m_c,\mu) + {\mathcal O}(m_c^2/p_T^2)\, ,
\label{cofact}
\end{equation}
where generally we pick the factorization scale $\mu$ to be of the 
order of $p_T$.
In this expression,  the convolution in the momentum fraction $z$ is 
denoted
by $\otimes$, and we have absorbed all information on the initial state
into $d\hat\sigma_{A+B\to g+X}$.
If we also assume NRQCD factorization, we have in addition to 
(\ref{cofact}),
\begin{equation}
d\sigma_{A+B\to H+X}(p_T) = \sum_n d\hat\sigma_{A+B\to 
c\bar{c}[n]+X}(p_T)\,
\langle {\mathcal  O}^H_n\rangle\, ,
\label{nrfact}
\end{equation}
where the ${\mathcal O}^{H}_n$ are NRQCD operators, classified by 
powers of
relative velocity and characterized by the various rotational and color
transformation properties of the $c\bar{c}$ state $[n]$.
Assuming both (\ref{cofact}) and (\ref{nrfact})
to hold, we conclude that the gluon fragmentation function is
related to the NRQCD matrix element by \cite{braaten96}
\begin{equation}
D_{H/g}(z,m_c,\mu) = \sum_n d_{g\to c\bar{c}[n]}(z,\mu,m_c) \, \langle 
{\mathcal O}^H_n\rangle\, ,
\label{combofact}
\end{equation}
where $d_{g\to c\bar{c}[n]}(z,\mu,m_c)$ describes the evolution of an 
off-shell
gluon into a quark pair in state $[n]$, including logarithms of 
$\mu/m_c$.

In the following, we will study the fragmentation function itself,
concentrating on infrared divergences at NNLO.  First, however, we
make some observations concerning the gauge transformation properties 
of NRQCD color octet
matrix elements.

\section{Gauge Invariance and Wilson Lines}

Production operators for state $H$ were introduced
in Ref.\ \cite{bodwin94} in the form
\ba
{\mathcal O}^H_n(0)
=
\chi^\dagger{\mathcal K}_n\psi(0)\, \left(a^\dagger_Ha_H\right)\,
\psi^\dagger{\mathcal K}'_n\chi(0)\, ,
\label{Ondef1}
\ea
where $a^\dagger_H$ is the creation operator for state $H$,
and where
${\mathcal K}_n$ and ${\mathcal K}'_n$ involve products of color
and spin matrices, and at
higher dimensions of covaraint derivatives.
Although the heavy (anti)quark fields ($\chi$) $\psi$
are all at the same space-time point (here $0$),
the operator ${\mathcal  O}^H_n$ is not truly local,
because the operator $a_H$ creates particle $H$
for out states, in the far future.  In particular, operator-valued gauge
transformations do not commute with the product
$a^\dagger_Ha_H$ in general.

A consequence of nonlocality is that
the right-hand  side of Eq.\ (\ref{Ondef1}) is
not gauge invariant in perturbation theory unless the individual
product $\psi^\dagger{\mathcal K}'_n \chi(0)$
and $\chi^\dagger{\mathcal K}_n\psi(0)$ are separately invariant.  This 
is the
case when the ${\mathcal K}_n$'s specify color singlets,
but not when they specify color octets.  A related issue
arises in the field-theoretic definitions of fragmentation
functions \cite{collins82}, such as $D_{H/g}$ in Eq.\ (\ref{cofact}) 
above.
    It is resolved by supplementing the
bi-local products of fields by nonabelian phase operators,
or Wilson lines: $\Phi_l [x,A] = \exp \left[-ig\int_0^\infty
d\lambda\,  l \cdot A(x+\lambda l )\right]$.
In contrast to parton distributions, moments of
fragmentation functions do not result in expectation values of local
operators \cite{mueller78}, and the corresponding Wilson lines
are not guaranteed to cancel identically.
For a color octet combination,
the gauge field $A^\mu$ is given in adjoint representation,
just as for a gluon fragmentation function, for which the operator
$\Phi_l (x)_{ab}$ multiplies the gauge covariant
field strength $F^{\mu\nu}_b$.
To be  definite,
and for convenience in relating
NRQCD operators to the  gluon fragmentation function,
we will choose $l $ as a lightlike vector in
the minus directon: $l^\mu =\delta_{\mu -}$.

Our gauge invariant redefinition of production operators
in octet representation is now found by the replacement
(which we refer to as a gauge completion),
\ba
{\mathcal O}^H_n(0)
\to
\chi^\dagger{\mathcal K}_{n,c}\psi(0)\, \Phi_l^\dagger 
[0,A]_{cb}\left(a^\dagger_Ha_H\right)\,
\Phi_l [0,A]_{ba}\, \chi^\dagger {\mathcal K}'_{n,a}\psi(0)\, ,
\label{replace}
\ea
where we have exhibited the color indices of the octets.
In perturbation theory, the Wilson lines generate
propagators of the form $i/(l \cdot k+i\epsilon)$
when they carry gluon momentum $k$, and the gluons
couple to the Wilson line at vertices $g_sl ^\mu C_{abc}$, with 
$C_{abc}$
structure constants.  These are precisely the same as
the propagators and vertices of the eikonal
approximation for the emission and absorption of soft gluons by
an  energetic gluon of momentum $l ^\mu$.  We will use this
correspondence below.

It is worth noting that the replacement (\ref{replace})
isn't really necessary if gauge dependence in the matrix elements
and coefficient functions is
infrared safe.  This appears to be an implicit assumption
in the discussion of Ref.\ \cite{bodwin94}.  As we  shall see,
however, although gauge dependence is certainly
free of infrared singularities at  next-to-leading
order, at NNLO this is no longer the case.

\section{Eikonal Approximation for $g\to H$}

For this discussion, we study the
perturbative expansion of the fragmentation
function for $g\to c\bar{c}[{n_0}]+X$ with
$c\bar{c}[{n_0}]= c(P/2+q)\bar c(P/2-q)$,
a pair with total momentum $P$
and relative momentum $q$, always projected onto
a color singlet state.  We can, of course, further project
various spin and orbital angular  momentum states $[n]$ as
in Eq.\ (\ref{combofact}).

Beyond lowest order, the computation of these diagrams
tests NRQCD factorization, Eq.\ (\ref{combofact}), which is
verified to the extent that we can absorb, that is match, any infrared
divergences into the perturbative expansion of NRQCD
matrix elements ${\mathcal O}^H_n$ \cite{braaten97b}.
We will be interested in gluons whose energies
are much below $m_c$.  For NRQCD factorization to hold
these soft gluons must factorize from
the higher energy radiation that
generates logarithms of $p_T/m_c$.
\footnote{We will show elsewhere that this is indeed the case.}

Because we need only the infrared structure of the diagrams
from  gluons of very low momentum, we may suppress
overall color and combinatorial factors, as well as momentum
factors that depend only on the scale of $m_c$, including
the LO fragmentation of the parent gluon, which is off-shell by
$\mu^2 \ge 4m_c^2$, into the quark pair.
\begin{figure}[h]
\centerline{\epsfxsize=5.5cm \epsffile{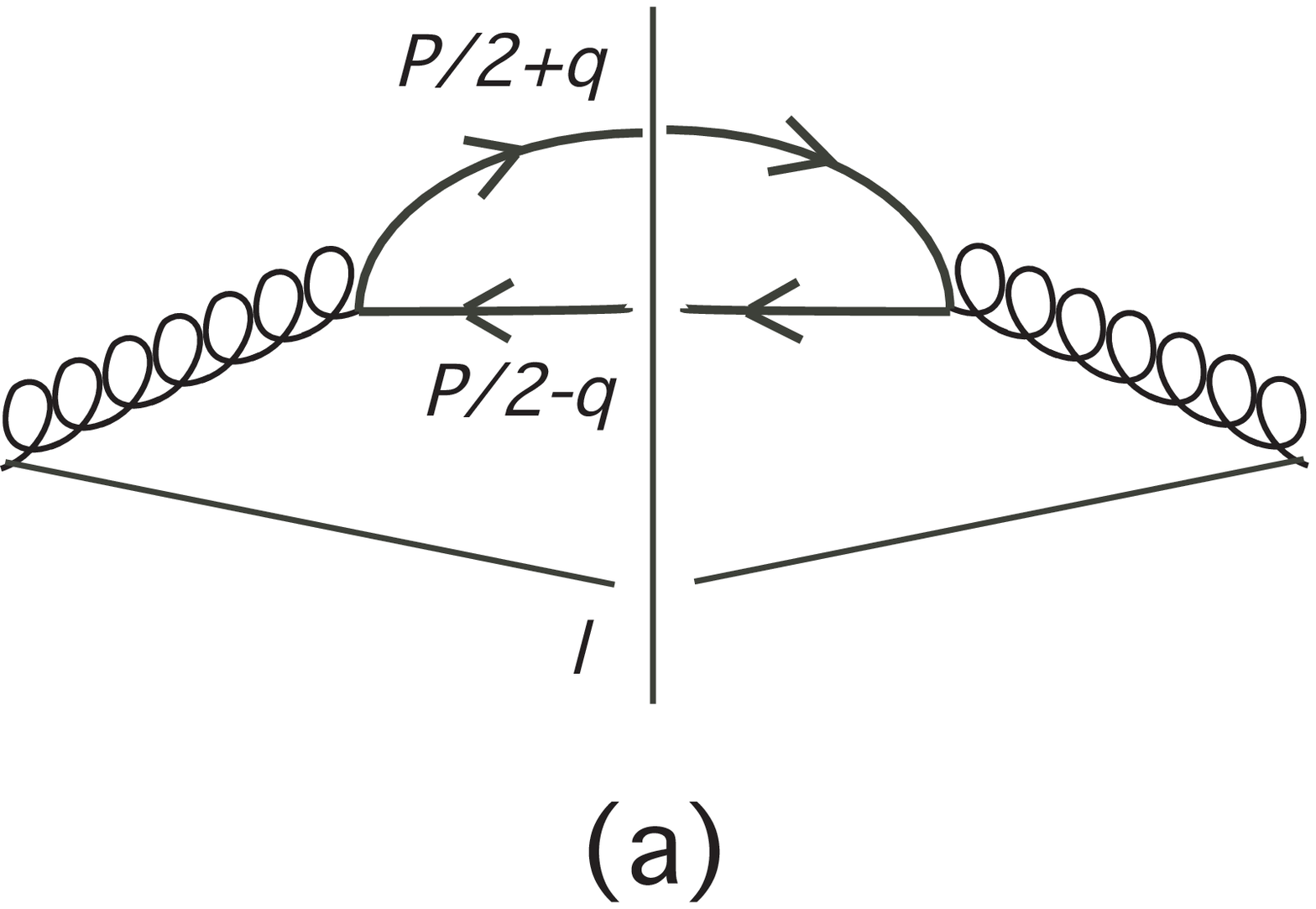} \quad 
\epsfxsize=5cm \epsffile{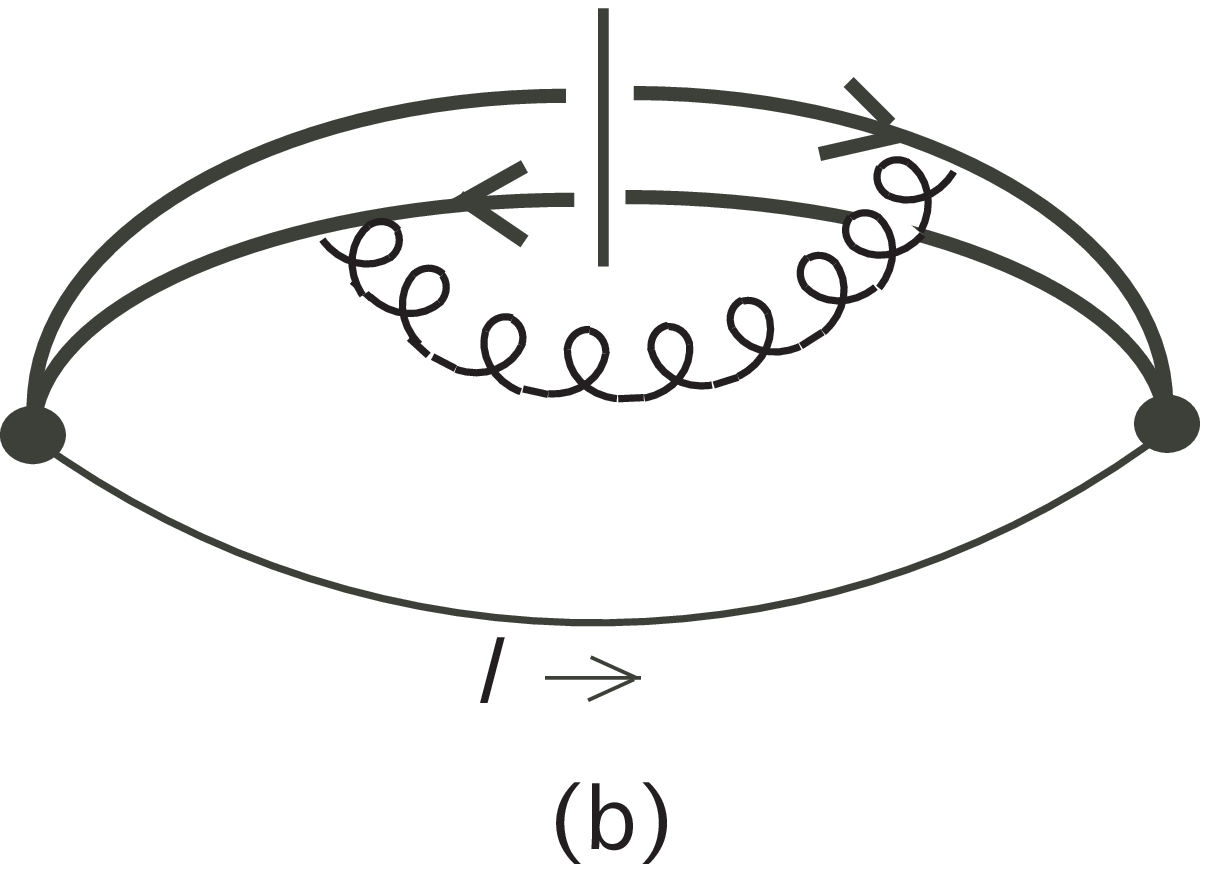}\quad \epsfxsize=5cm 
\epsffile{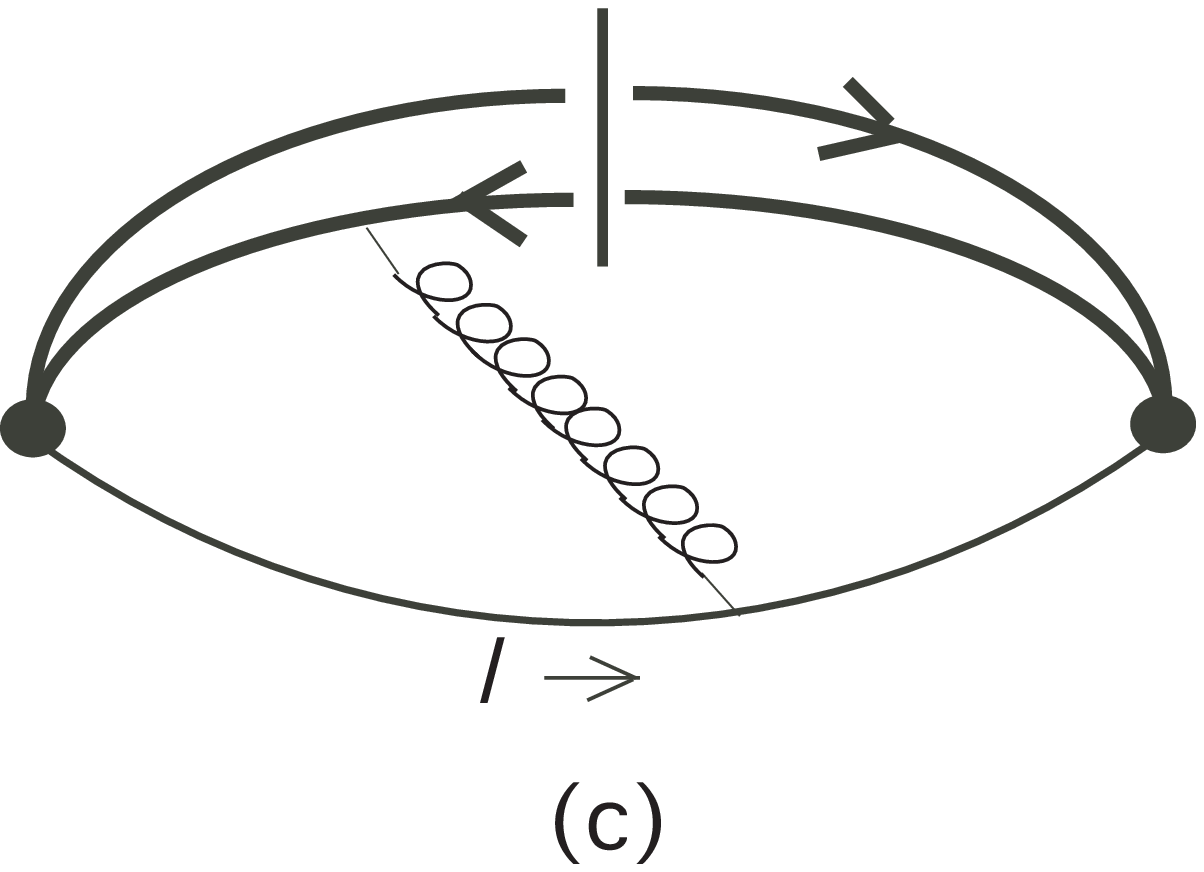}}
\label{fig1}
\caption{a) Lowest-order fragmentation function for $g\to c\bar{c}$.  
There are no interactions on the eikonal quark pair or the Wilson line 
that corresponds to
an eikonal gluon of four-velocity $l$.   b,c)
Representative NLO contributions to $g\to c\bar{c}$ fragmentation in 
eikonal approximation.
In these figures the parent gluon is contracted to a point, represented
by the dark circle, because it is off-shell by
order $m_c$}.
\end{figure}

Figure 1a shows the lowest order diagram $g\to c\bar c$,
and Figs.\ 1b,c show typical contributions
with single soft gluon emission in
the fragmentation function, all in cut diagram notation.
Both Fig.\ 1b and 1c represent a class of
four diagrams, in which gluons are attached to one of the two
lines of the quark pair in the amplitude and its complex conjugate.
In each diagram we may treat the quarks in eikonal approximation,
in effect replacing them with time-like Wilson lines, whose
propagators and vertices are given by $i/(\beta\cdot k+i\epsilon)$
and $-ig_sT_a\beta^\mu$, respectively, for quarks, with $T_a$ a
color generator in fundamental representation and
$\beta^\mu$ the time-like quark four-velocity.  The
corresponding approximation for antiquarks simply changes the
sign at the vertex.

The vertical lines in Fig.\ 1b and 1c represent the quark pair final 
state,
including a projection onto
a color singlet configuration.   The full fragmentation
function is found by cutting the remaining lines in
all possible ways.
When the gluon is cut in Fig.\ 1b, in particular,
the resulting diagrams have an uncancelled infrared
pole in dimensional regularization,
  beginning at order $v^2\sim \vec{q}\, {}^2/m_c^2$
when the gluon momentum $k\to 0$.
This gluonic contribution, however, which involves only interactions
between the pair, may be matched
straightforwardly to a color octet matrix element,
with which it shares an identical topology \cite{braaten96,braaten97b}.

The divergent part of Fig.\ 1b may be isolated by evaluating
the four relevant diagrams in eikonal approximation and then
expanding them in $v=q/m_c$.  Alternately, we may expand
the momentum-space integrand,
isolating the $v^2$ term  before the integration.
In $D=4-2\varepsilon$
dimensions, and in the quark pair rest frame, the latter approach leads
readily to an
expression in terms of the total and relative momenta
$P$  ($P^2=m_H^2$) and $q$, respectively,
\begin{eqnarray}
\Sigma^{(1b)}(P,q)~
&=&~ 16\, g^2\mu^{2\varepsilon}\, \int \frac{d^D 
k}{(2\pi)^{D-1}}~\delta(k^2)\,
    [q_\nu (P\cdot k)-(q\cdot k) P_\nu] \nonumber\\
&& \hspace{30mm} \times
\ [q^\nu (P\cdot k) - (q\cdot k) P^\nu]
\frac{1}{[(P\cdot k)^2]^2} \nonumber\\
&=& {16\over 3}\, {\alpha_s\over \pi}\, {\vec{q}\, {}^2\over P^2}\, 
{1\over -\varepsilon} + \dots\, ,
\label{nlopt}
\end{eqnarray}
where we have exhibited the infrared pole, regularized for 
$\varepsilon<0$.
This infrared pole, along with an appropriate
color trace, appears at NLO as a multiplicative factor times
the lowest-order fragmentation function,
as found, for example, in Refs.\
\cite{braaten97,petrelli98}.

In  the integral, the terms in square brackets correspond
to the lowest-order vertices for the operator
$q^\alpha F_{\alpha\beta}P^\beta$, with $F$ the field
strength.  In the quarkonium rest frame, this corresponds
to an electric dipole transition from
octet to singlet color states \cite{braaten96}.
An advantage of this expansion is that, because
the field strength decouples from scalar polarized
lines, all  singularities associated with gluon momenta
collinear to $l^\mu$ vanish
on a diagram-by-diagram basis, and double poles
in $\varepsilon$ are entirely absent in the calculation.
By the same token, the integral in (\ref{nlopt}) is
gauge invariant in the polarization sum for
the final-state gluon of Fig.\ 1b.

Finally, in Fig,\ 1c,  the soft gluon connects the pair to the
Wilson line.  In covariant gauges, this diagram
has double and single infrared/collinear poles in dimensional 
regularization.
These poles, however, cancel against
analogous contributions from the corresponding virtual diagram.
Imaginary parts of the virtual diagrams, of course, cancel in
the fragmentation function, which is real.
In practical terms, the contribution of Fig.\ 1c plus its virtual
counterpart contributes solely to the perturbative
factor $d_{g\to c\bar{c}[n]}(z,\mu,m_c)$ in Eq.\ (\ref{combofact}).

We note that the infrared behavior of all of the
diagrams of Fig.\ 1 is common between the fragmentation function
and a generic cross section in which a high-$p_T$ gluon
of momentum $l $ recoils against the heavy quark pair.
Indeed, the same cancellation mechanism for the infrared divergences of
Fig.\ 1c is referred to specifically in Ref.\ \cite{bodwin94}
as an essential step in the justification of NRQCD
factorization for production cross sections.
To the extent that all the infrared poles of diagrams that do not share 
the topology
of NRQCD matrix elements cancel, the matching of
perturbative infrared poles with those of the effective
theory is ensured.  The result of this cancellation
is referred to as topological factorization \cite{braaten96}.
The same considerations that lead to the inclusion of
nonabelian phases to the fragmentation function, however,
suggest that the cancellation of non-factored soft gluons
may be nontrivial, and should be checked beyond NLO.
Of course, a full NNLO calculation of either cross sections
or fragmentation functions would be daunting.
Fortunately, the analysis of the relevant infrared behavior at NNLO
requires only the eikonal approximation,
and is therefore a much more manageable, although still extensive, task,
to which we now turn.

\section{Fragmentation Function at NNLO}

We can readily generalize our NLO discussion to NNLO.
Once again, because we are interested in gluons whose
energies are much less that $m_c$, we can adopt
the eikonal approximation, and neglect all
dynamics at the scale of the quark masses.
In NRQCD, these approximations are relevant to
gluons of order $m_cv$, with $v$ the relative velocity,
but of course in the perturbative calculation the gluon momentum
goes to zero.
We note that for octet-singlet transitions at NNLO, we need not include
virtual gluon exchange between the quark and antiquark,
except for diagrams that are already topologically
factorized.  As a result, we will not need not consider
violations of the eikonal approximation from momentum
regions at or below the scale $m_cv^2$, or
the resulting $1/v$ singular behavior.

Representative diagrams for the fragmentation function
are shown in Fig.\ 2.   The full infrared
fragmentation function is again generated by taking all
allowed cuts of the remaining lines of each such diagram.

We are concerned only with diagrams that connect octet to
singlet quark states, and which are not topologically
factorized, since these are the potential sources of
nonfactoring behavior in both the fragmentation function and
related cross sections.
We recall once  more that the familiar argument  for NRQCD factorization
is based on the conjecture that all infrared
regions in these diagrams cancel after this limited
sum over cuts \cite{bodwin94}.
In fact, we shall see that this is
the case at NNLO only if we employ the
gauge-completed definitions for NRQCD
matrix elements, as in Eq.\ (\ref{replace}) above.
\begin{figure}[h]
\centerline{\epsfxsize=5cm \epsffile{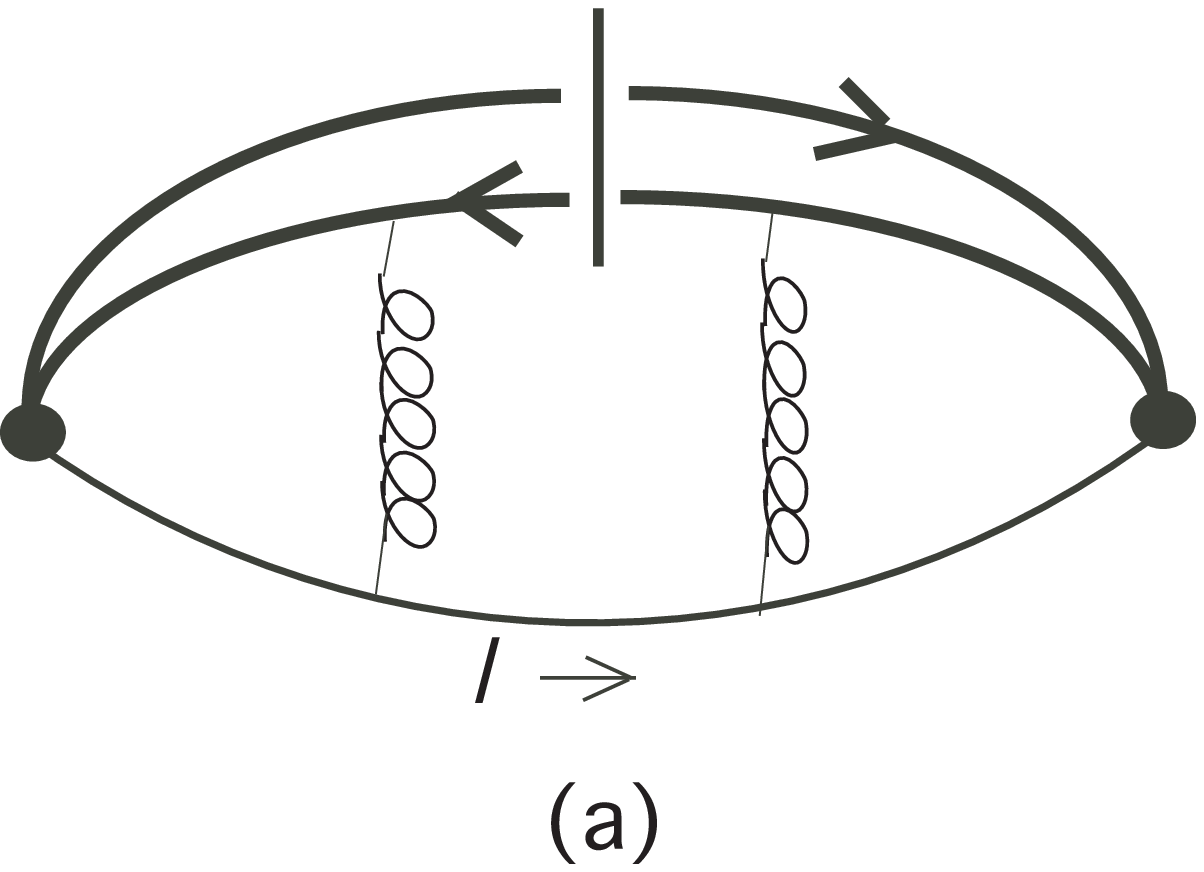} \quad 
\epsfxsize=5cm
\epsffile{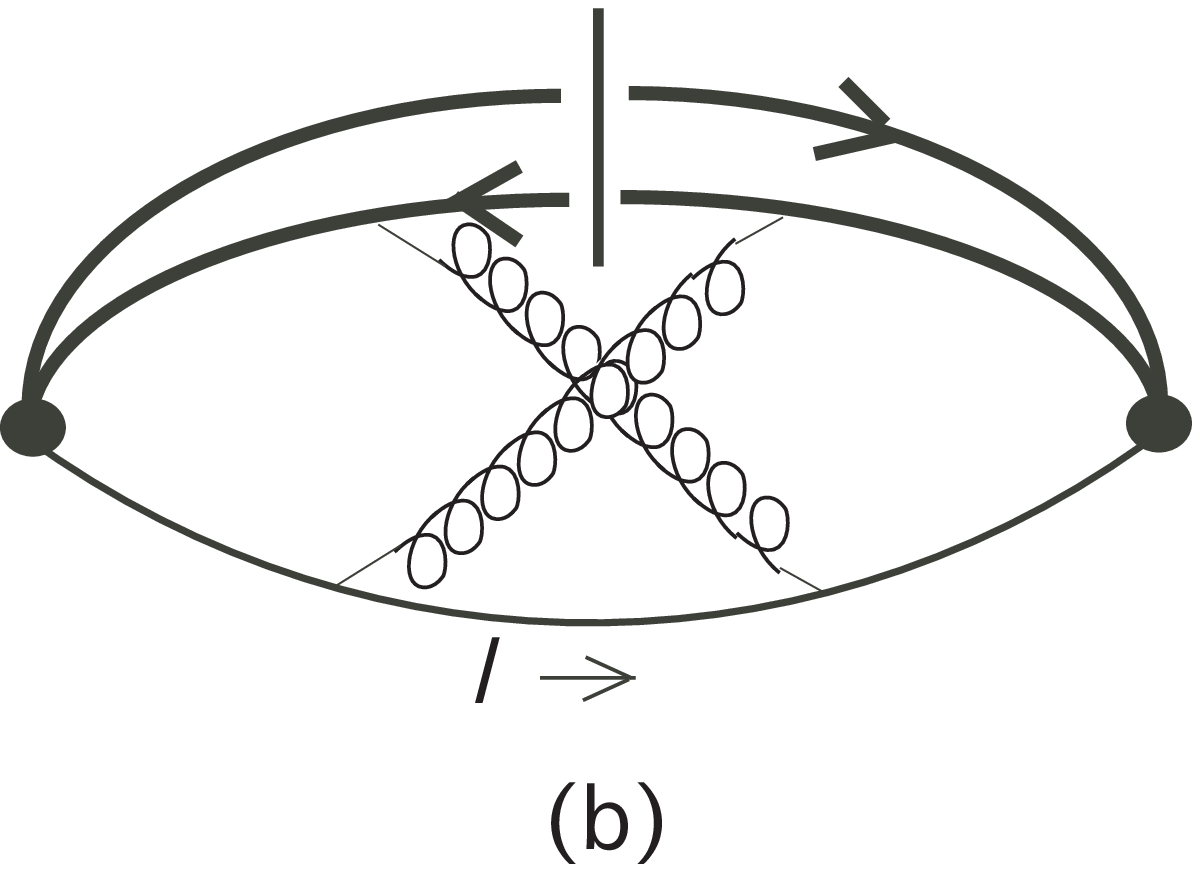}\quad \epsfxsize=5cm \epsffile{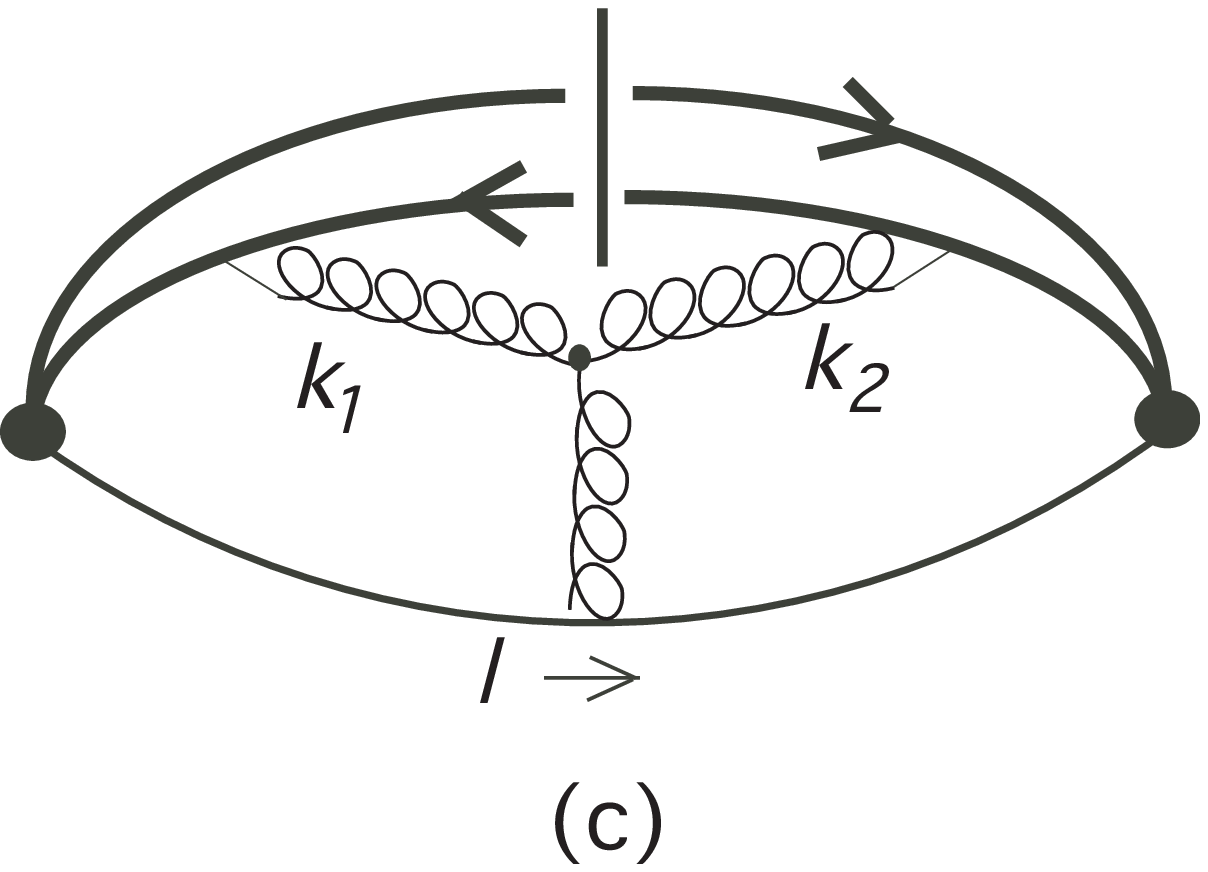}}
\label{fig2}
\caption{Representative NNLO contributions to $g\to c\bar{c}$ 
fragmentation in eikonal approximation.}
\end{figure}

As at NLO, we will concern ourselves here
with infrared behavior at order $v^2$, which
may be computed by evaluating each
diagram in eikonal approximation and then
expanding their sum in $q/P$, or by expanding the diagrams first, as
in Eq.\ (\ref{nlopt}).
Noting that the two
approaches give the same answer,
we summarize the results here, and
provide details of the calculations
elsewhere \cite{nayakip}.

The individual classes of
diagrams in Fig.\ 2a and 2b, for which
two gluons are exchanged between the
quarks and the Wilson line, satisfy the infrared
cancellation conjecture of Ref.\ \cite{bodwin94}, by summing over the
possible cuts and connections to quark and
antiquark lines, as do diagrams that have three gluon-eikonal
vertices on the quark pair and one on the  Wilson line.  For the class 
of diagrams
related to Fig.\ 2c, however, with a three-gluon interaction, this 
cancellation
fails.  Expanding again to second order
in the relative momentum $q$, the full
contribution from Fig.\ 2c, found by cutting the gluon line
$k_1$ and the Wilson line  can be written
by analogy to Eq.\ (\ref{nlopt}) as
\begin{eqnarray}
&&\Sigma^{(2c)}(P,q,l)~
=~ -16i\, g^4\mu^{4\varepsilon}\, \int \frac{d^D k_1}{(2\pi)^D} 
\frac{d^D
k_2}{(2\pi)^D}~2\pi~\delta(k_1^2) \;
l^\lambda\, V_{\nu\mu\lambda} [k_1,k_2] \nonumber\\
&& \hspace{40mm} \times
\ [q^\mu (P\cdot k_1) - (q\cdot k_1) P^\mu] \,
  ~ [q^\nu (P\cdot k_1)-(q\cdot k_2) P^\nu]~  \nonumber \\
&& \hspace{40mm} \times \frac{1}{[P\cdot k_1 +i\epsilon]^2~
[P\cdot k_2 - i\epsilon ]^2} \nonumber\\
&& \hspace{40mm} \times \frac{1}{[k_2^2 - i \epsilon]~[(k_2- k_1)^2 - 
i\epsilon]~
  [l\cdot (k_1 - k_2) - i\epsilon]}
\, ,
\label{nnlopt}
\end{eqnarray}
where $V_{\nu\mu\lambda} [k_1,k_2]$ represents the
momentum part of the
three-gluon coupling.  As in Eq.\ (\ref{nlopt}), we have suppressed
color factors and momentum dependence
at the scale $m_c$.

As observed above, the field-strength vertices eliminate
collinear singularities on a diagram-by-diagram basis.
The leading singularities in (\ref{nnlopt}) and related
diagrams are therefore never worse than $1/\varepsilon^2$.
Summing over all such contributions, however, we find
a noncancelling real infrared pole in the
fragmentation function, which may be written
in invariant form as
\begin{equation}
\Sigma^{(2)}(P,q,l)~=
~\alpha_s^2~\frac{4}{3 \varepsilon}~[\frac{(P\cdot
q)^2}{P^4}~-~\frac{q^2}{P^2}]\, .
\label{gn}
\end{equation}
In the rest frame of heavy-quarkonium ($\vec{P}=0$),
this becomes simply
\begin{equation}
\Sigma(P,q,l)~=~\alpha_s^2~\frac{4}{3 \varepsilon}~\frac{\vec{q}\, 
{}^2}{4m_c^2}
~=~\alpha_s^2~\frac{1}{3 \varepsilon}~\frac{\vec{v}\, {}^2}{4}\, ,
\label{gn1}
\end{equation}
where $\vec{v}$ is the relative velocity of the heavy quark pair.
We will give the extension
of this result to all powers in $v$ elsewhere \cite{nayakip}.

Eq.\ (\ref{gn1}) shows explicitly the breakdown of the simplest 
topological
factorization of infrared dependence at NNLO.
Its presence implies that
  infrared poles would appear in coefficient functions at NNLO
and beyond when the factorization is carried out with octet
NRQCD matrix elements defined in the conventional manner,
Eq.\ (\ref{Ondef1}).  On the other hand, when defined according
to its gauge-completed form (\ref{replace}), each octet
NRQCD matrix element itself generates precisely the
same pole terms given in (\ref{gn1})  above.
Thus, at least at NNLO and  to order $v^2$,
NRQCD factorization can accomodate these corrections.
We note, however, that at this order
the NNLO correction is independent of the direction $l$ of
the Wilson line, while   Lorentz invariance alone would
seem to allow a correction of the form $(q\cdot l)^2/P^2$.
The presence of poles with this coefficient could indicate
problems with factorization, that is, with matching
the poles of the fragmentation function to those of the
perturbative cross section.  These could arise when the
integration over $q$ is not rotationally invariant, as might
happen for polarized final states, and/or final states with
gluons of energies of order $m_c$ in the rest frame of
$H$, since their directions are arbitrary.   We are unable at this 
stage to rule out
the occurance of such terms at higher orders in
the strong coupling, or in higher powers of $v^2$.

\section{Summary and Future Studies}

We have discussed the gauge invariance properties of
color octet matrix elements in NRQCD as they
appear in fragmentation functions.  Gauge invariance
and factorization require that they include
nonabelian phase operators to match otherwise nonfactoring
  infrared divergences beginning at NNLO.

We have shown that at NNLO there are infrared (not collinear)
poles from fragmentation diagrams that
are not topologically equivalent to conventional NRQCD operators.
We have observed that rotational invariance
in the quarkonium rest frame is
an important consistency
condition for the possibility of NRQCD factorization
in terms of the gauge-completion of NRQCD operators.
Should rotational invariance fail at higher orders,
we anticipate that it will be necessary to complement
the NRQCD classification of nonperturbative
parameters for quarkonium production with
matrix elements of additional  operators,
involving nonlocal products of the field strengths encountered above.

Finally, although our explicit
calculations are carried out in the eikonal
approximation for gluons of energies far below
$m_c$, it is not difficult to verify that
these gluons do not interfere with the generation of standard
evolution for the $g\to H$ fragmentation
functions.  The arguments for this result
will be given elsewhere, along with details
of the calculations described above.

\subsection*{Acknowledgements}

This work was supported in part
by the National Science Foundation, grants PHY-0071027, PHY-0098527 and 
PHY-0354776,
and by the Department of Energy, grant DE-FG02-87ER40371.
We thank Geoff Bodwin for
emphasizing the importance of this topic,
and for many useful discussions.

\end{document}